\documentclass{IEEE-Con-Sys-mag}

\jvol{XX}
\jnum{XX}
\paper{8}
\jmonth{June}
\jname{IEEE CONTROL SYSTEMS}
\pubyear{2020}

\usepackage{enumerate}
\usepackage{float}
\usepackage{graphicx} 
\usepackage{subcaption}
\usepackage{xcolor}

\def\lab{\label}
\def\bul{\noindent $\bullet\;\;$}

\def\begenu{\begin{enumerate}}
\def\endenu{\end{enumerate}}
\def\begequ{\begin{equation}}
\def\endequ{\end{equation}}

\def\begmat#1{\begin{bmatrix}#1\end{bmatrix}}
\def\begalis#1{\begin{align*}{#1}\end{align*}}

\def\rea{\mathbb{R}}
\def\calh{{\cal H}}

\def\sign{\mbox{sign}}
\def\et{\varepsilon_t}
\def\diag{\mbox{diag}}

\def\CEP{{\it Control Engineering Practice}}

\def\IJRNLC{{\it Int. J. on Robust and Nonlinear Control}}
\def\TAC{{\it IEEE Trans. Automatic Control}}
\def\TIE{{\it IEEE Trans. Industrial Electronics}}

\def\IJC{{\it International Journal of Control}}

\def\AUT{{\it Automatica}}

\begin{document}

\title{Some Reflections on Sliding Mode Designs in Control Systems\stitle{An Example of Adaptive Tracking Control for  Simple Mechanical Systems With Friction Without Measurement of Velocity}}

\author{{R}OMEO ORTEGA, LEYAN FANG and JOSE GUADALUPE ROMERO}

\maketitle

\section{Introduction}

\textit{Automatic control} emerged as a multidisciplinary discipline in the mid 20th century, see {\em e.g.}, \cite{ASTKUM} for an illuminating description of its history. The International Federation of Automatic Control (IFAC) was formed in 1956, the first IFAC World Congress was held in Moscow in 1960, and the journal \AUT$\;$  appeared in 1962. Control is a key enabler of several technological fields, including electrical, mechanical and chemical. Respect for mathematical rigor has been a hallmark of control systems research, whose development has been mainly realized by engineers and mathematicians, with the former been essentially motivated by the need to understand, design and maintain man-made engineered systems. On the other hand, mathematicians were more concerned by the construction of a solid mathematical basis for control theory. One consequence of the ever increasing involvement of mathematicians in the field was that the parity between theory and practice was breached. Pure theory seized attention to a significant extent to the point that there appeared a ``gap" between theory and practice \cite{BER}.  

Our motivation in writing this paper is to contribute towards the reversal of the situation described above. The authors, as engineers, have developed their research in automatic control, always trying to preserve a contact with real world applications---avoiding the proposals motivated by pure mathematical considerations. We were skeptical of the study of nonlinear control of systems with special mathematical structures, {\em e.g.}, triangular forms \cite{KRSKANKOKbook}, which were not motivated by physical considerations, but only by our mathematical ability to deal with them. In the present contribution, we bring to the readers attention the fact that a similar situation prevails in most of the research reported on  {\em sliding mode} (SM). SM is a nonlinear control method belonging to the class of variable structure control systems introduced by Emelyanov in the 60s \cite{EMEbook}---see, {\em e.g.}, \cite{FERINCCUCbook,UTKbook} for details and \cite{POZORL} for a vivid desciption of its history. In this method, the state of the system is enforced to be constrained in a subspace called a sliding surface, where the state of the system evolves according to the desired dynamics. To achieve the SM, discontinuous---relay induced---{\em high gain} feedback is employed to enforce the state to the sliding surface. However, such input induces undesirable {\em chattering} phenomena a problem that has been studied for over 50 years---with a modest (41 years old) contribution of the first author \cite{ESPORTESP}, which was motivated by its appearance in experimental facilities \cite{ORT}---see \cite{ARAetal} for some compelling experimental evidence of its presence in SM designs. In spite of many efforts---see for instance \cite{LABEFU}---chattering seems to be intrinsic, hence unavoidable, to the SM design.

The remainder of our paper is organized as follows. First, we present the aforementioned comments on SM design. Second, we use a specific example to illustrate our previous considerations. Towards this end, we compare the performance of two adaptive tracking controllers for a simple one degree of freedom (DoF) mechanical system with unknown parameters and static and Coulomb friction---that do not rely on the measurement of velocity. Both adaptive controllers are speed observer-based, they are designed following the Immersion and Invariance (I\&I) methodology \cite{ASTKARORTbook} and the theory of second order sliding modes \cite{LEV}. Finally, we wrap-up the paper with some concluding remarks and present, in Appendix \ref{appa}, some considerations regarding the tuning procedure of the I\&I design.

\begin{summary}
\summaryinitial{T}he objective of this note is to share some reflections of the authors regarding the use of sliding mode designs in control systems. We believe the abundant, and ever increasing, appearance of this kind of works on our scientific publications deserves some critical evaluation of their actual role, relevance and pertinence. First, we discuss the procedure followed by most of these designs---illustrated with examples from the literature. Second, we bring to the readers attention several aspects of the control problem, central in classical designs, which are disregarded in the sliding mode literature. Finally, to illustrate with an specific example our previous considerations, we compare the performance of two adaptive tracking controllers for a simple one degree of freedom mechanical systems with unknown parameters and static and Coulomb friction---that do not rely on the measurement of velocity.
\end{summary}

\section{Comments on Some SM Designs and Three Examples}
\label{sec2}

\textit{Caveat:} The description of SM given in this section pertains only to some papers---including the ones cited here---and it is not claimed to be applicable {\em to all} SM designs. Actually, there is an effort by some researchers to try to develop and alternative basis for SM designs, exploiting fundamental physical properties like passivity, see {\em e.g.},  \cite{SAKFUJICH}---which is a very welcomed initiative.\\

In this section we first describe the three major steps that some SM designs follow to achieve their objective. Then, we illustrate this point with some specific examples taken from the literature. 

\subsection{Three basic steps for SM designs}
\lab{subsec21}
\bul The first step in some SM designs is to "rewrite" the system dynamics in a form that retains a "good" part of the system perturbed by an {\em additive} undesirable "disturbance".\footnote{SM is not the only technique that adopts this practically questionable perspective for control design, a similar procedure is adopted in the {\em active disturbance rejection} methods \cite{SHEetal}, with an extreme position taken in the {\em model-free control} of \cite{FLIJOI}, that claims that all systems can be described by an "ultra-model" consisting of a perturbed $\nu$-order integrator \cite[Equation (1)]{FLIJOI}---with $\nu=1,2$.}  This rewriting is done disregarding all structural properties of the system that in other control designs are exploited to provide a sensible solution to the problem. 

\bul The second step is to impose the highly disturbing assumption that the {\em state is bounded}---and sometimes even the derivative of the state (see comment {\bf C1} in Subsection 3.2.2 below). The "argument" invoked to impose this assumption is that we are dealing with physical systems, hence all physical variables are bounded, however, the simulations which are shown in SM papers never include saturations in the state variables. It is difficult to accept this kind of argument pertaining to one of the key properties that need {\em to be proven} on regular control systems designs: that all signals are bounded!

\bul In the third, and final, step the SM design incorporates a mechanism to inject high gain to the loop, usually including relay functions and/or terms with fractional powers. The high-gain injection, of course, amplifies the noise, while the relay action induces the aforementioned undesired chattering phenomenon.\footnote{It is quite revealing to note that the plot of the control signal is very rarely shown in simulation of SM control---when this is done we see a physically unrealizable black blotch \cite[Figure 7]{JIAetal}.}  To be able to prove that the high-gain injection actually dominates the "undesirable" terms in the dynamics (treated as "disturbances") it is, furthermore, assumed that their constitutive functions are {\em globally Lipschitz}---ruling out all physical examples of practical interest.

The overall design of the observer and controller is usually extremely involved, with many tuning gains of imprecise role. The issue of including in the design tuning gains whose effect on the performance is imprecise plagues most SM-based schemes, and makes very difficult the task of properly tuning the free gains of the design---a key step to improve performance. As a token of illustration we repeat here \cite[Theorem 1]{ESTetal}:\\

{\bf Theorem 1.} {\em Suppose Assumptions 1-3 are satisfied. Consider the control law (13) with (16) in a closed-loop system (6). For an {\em appropriate selection} of two tuning gains, there exist {\em sufficiently large} values for two more tuning gains that ensures the tracking error goes to zero in {\em some finite time}}.\\

We believe this statement provides little help to the potential user that needs to tune the controller reported in \cite{ESTetal}! It is worth mentioning that---in the spirit of the step one described above---in this paper the highly nonlinear hydraulic actuator dynamics  \cite[Equations (1)-(5)]{ESTetal} is "approximated" by a {\em linear} system perturbed by two additive functions  \cite[Equation (6)]{ESTetal}, assumed Lipschitz and, of course, verifying that the state, and its derivative are  bounded.

Before closing this subsection we would like to bring to the readers attention the following two facts:
\begenu[{\bf F1}]
\item In the latest developments on this area, namely the use of high-order SM \cite{LEV}, contrary to conventional wisdom in control theory \cite{OGAbook}, it is suggested that it is possible to {\em differentiate} the signals as many times as desired---see also \cite{CRUMOR} for similar developments.
\item Even at a purely mathematical level, there is some discussion on the treatment of differential equations with discontinuous right-hand sides. The  simplest regularization, widely adopted in the SM literature, is Filipov's method. However, this approach was severely criticized, even by the founder of SM theory in 
\cite{UTKbook}.  As indicated in \cite[Subsection 2.1.2]{POLbook}: "It is widely recognized that  Filippov's regularization does not describe correctly some real control systems with discontinuous models, and a proper modification of Filipov's definition---not always clarified by the authors---is usually required." The interested reader is referred to \cite[Subsection 2.1.2]{POLbook} for a detailed discussion on this topic.
\endenu

\subsection{Example of regulation of mechanical system}
\lab{subsec22}
%
An example of SM design is reported in \cite{OVAetal}, where the classical problem of position-feedback regulation of underactuated $n$-degrees of freedom (DoF) mechanical systems is studied---problem which, by the way, was solved 10 years ago in  \cite{ROMORTSAR}. In   \cite{OVAetal} it is proposed to rewrite the very well-known and highly structured Euler-Lagrange equations \cite[equation (1)]{OVAetal} into a chain of integrators ``perturbed" by some functions \cite[equation (4)]{OVAetal} of the form\footnote{We disregard the presence in \cite[equation (1)]{OVAetal} of a torque external disturbance that is later dominated with a high gain design.}
\begequ
\lab{sys}
\begmat{\dot x_u \\ \dot x_v \\\dot x_a \\ \dot x_b}=\begmat{x_v \\ f_u(x)+g_u(x_u)\tau \\ x_b \\ f_a(x)+g_a(x)\tau},
\endequ
where $x_u(t) \in \rea^{n_u}$ are the unactuated positions and $x_a(t) \in \rea^{n_a}$ the actuated ones, with $n=n_u+n_a$, both of them assumed measurable. The motivation to consider this contrived structure is that a sliding manifold, which is {\em linear} with respect to the {\em full state} vector, can be imposed and a finite-time convergence high-order SM observer for the residual dynamics on the manifold can be designed. The second stage is the design of a SM controller based on the super-twisting algorithm. The overall design of the observer and controller is extremely involved, with many tuning gains of imprecise role. 

In order to ensure that the mechanical system can be re-written in the form \eqref{sys} a long list of {\em physically unmotivated} assumptions are imposed on the systems inertia matrix and the gravity vector \cite[Assumption 1]{OVAetal}. A particularly critical assumption is that the part of the gravity function dependent on the unactuated coordinates is {\em zero only at zero}. This assumption rules out almost all practical systems, {\em e.g.}, robots with rotational joints, pendular systems, whose objective is to lift the pendulum, etc. The example used to illustrate the theory has a hanging pendulum! On the other hand, the assumptions imposed on the inertia matrix imply that the $2 \times 2$ sub-block of the inertia matrix is {\em constant}. 

It is interesting to note that, under the assumptions made on the inertia matrix in \cite{OVAetal} the underactuated mechanical system can be globally stabilized with a simple {\em PID passivity-based control} \cite[Proposition 7.1]{ORTetalbookpid}. Moreover, the highly restrictive assumption on the gravity force imposed in  \cite{OVAetal} is {\em not required} for this basic design.
\subsection{Example of blood glucose regulation}
\lab{subsec23}
%
In \cite{FRAetal} a study on the highly ambitious project of regulation of the blood glucose levels in critically ill patients affected with Type 1 Diabetes Mellitus. The system to be controlled is modeled as \cite[Equation (1)]{FRAetal}
\begalis{ 
\dot x&=Ax+\begmat{-x_1 x_2+d(t)\\ 0 \\ u}\\
y & = x_1
}
with $x(t) \in \rea^3$ the {\em unmeasurable} state, $u(t) \in \rea$, $y(t) \in \rea_+$, the system input and output, respectively, $A \in \rea^{3 \times 3}$ and $d(t) \in \rea$ is an {\em exponentially decaying} disturbance. The control objective is to regulate the output around some constant desired value $y^\star >0$.

The first step of the SM design described above is achieved differentiating three times the regulation error $e:=y-y^\star$ to get  \cite[Equation (4)]{FRAetal}
$$
{d^3 \over dt^3}e= \phi(t,x)+v,
$$ 
where---assuming $x_1$ bounded away from zero---we defined\footnote{This operation is carried out in all the controllers proposed in  \cite{FRAetal}.} 
$$
u={1 \over p_3 x_1}v,
$$
with $p_3>0$, and $\phi(t,x)$ is a complicated nonlinear function of $x$ and $t$ given in \cite[Equation (5)]{FRAetal}. 

Following the second step of the SM design it is assumed that the ``disturbing" function $\phi(t,x)$, and its derivative, are {\em bounded} with {\em known} bounds for both functions. 

To complete the third step {\em five} different SM algorithms were proposed---with the intention of comparing their performances when applied in such a trivial example. All the discontinuous controllers associated to the five SM strategies depend on the bounds on the disturbance (and its derivative) described above and some tuning parameters. But no discussion is made on how to select these gains. Actually, in one of the theoretical results \cite[Theorem 5]{FRAetal} there are eight tuning gains $k_i>0,i=1,\dots,6$ and $L>0,H>0$, and it is simply stated: ``there exist constants $k_i>0$ and $L_m>0,H_m>0$ such that for all $L>L_m$ and $H>H_m$ the control objective is achieved". This seems to be of little help to select those eight gains.             
{
\subsection{Example of magnetic levitation systems}
\lab{subsec24}
%
A magnetic levitation ball system with no iron coil magneto-resistance is considered in  \cite{WANGetal}. The well-known, highly nonlinear, model describing its behavior is given in  \cite[Equation (1)]{WANGetal}.\footnote{This system has been extensively studied in the control theory literature, with many systematic design techniques successfully applied---see \cite{RODSIGORT,TORORT} for a comparison of some of them.}  The control objective is to ensure that the ball position tracks a desired trajectory $x_d(t) \in \rea$ in finite time.

The first step of the SM design described above starts by partially {\em linearizing} the model around some operation mode yielding 
\begalis{ 
\ddot x&=a_0x+ b_0u+f(i,x,u,d,t)\\
y & = x+n
}
with $x(t) \in \rea$ is the ball position, $u(t)$, $y(t) \in \rea$ are the system input and output respectively and $a_0>0$, $b_0>0$ are some {\em known} constants. The term $f$ is a highly complicated function of $(x,u,t)$, the coil current $i(t) \in \rea$ and an undefined function $d(t) \in \rea$, treated as a ``disturbance". 

Following the second step, it is assumed that $f$ is {\em bounded} by a {\em known constant}  \cite[Assumption 1]{WANGetal}.   

To complete the third step {\em three} different adaptive terminal SM algorithms were proposed---with the intention of comparing their performances when applied in such a trivial example. All the discontinuous controllers associated to the three SM strategies depend on the bounds on the disturbance described above and some tuning parameters. Moreover, as the complexity of the proposed SM algorithms increases, the number of tuning parameters grows from two to five and eventually to eight; however, discussion on how to select these parameters is very limited. It is interesting that the authors include some zero mean Gaussian noise in their simulations---that, apparently, is represented by the signal $n$ in the model above---and present the control signal  \cite[Figure 3]{WANGetal}. Interestingly, experimental results are also presented. 
}
  
\section{An Example of Adaptive Tracking Control}
\lab{sec3}
%
To illustrate some of the issues discussed above, in this section we present an specific example of the control of a mechanical system, for which solutions based on SM and I\&I theory have been reported.   
\subsection{Model of the system and problem formulation}
\lab{sec31}
%
The dynamics of a planar one DoF mechanical system with static and Coulomb frictions is given by
\begin{eqnarray}
\dot x_1 &=&x_2 \nonumber\\
\dot x_2&=&- \theta_1 x_2 -\theta_2 \tanh(\vartheta x_2) + u,
\label{sys1}
\end{eqnarray}
where  $x_1(t) \in \rea$ and $x_2(t) \in \rea$  are the generalized position and velocity, respectively, $u(t) \in \rea$ is the control input and $\theta_1>0$, $\theta_2>0$ and $\vartheta>0$ are constant coefficients. To simplify the notation we have lumped the motor inertia, which  is assumed known, into the control signal and the constants $\theta_1$ and $\theta_2$. The assumption of known motor inertia is done without loss of generality, since its incorporation as an additional unknown parameter in the adaptive estimators proposed below is  straightforward.  

The control objective is to design a globally convergent adaptive tracking regulator based on a speed observer for $x_2$, which is designed considering that only $x_1$ is {\em measurable}, $\vartheta$ is {\em known} and $\theta_1$ and $\theta_2$ are {\it unknown}.\footnote{The assumption that $\vartheta$ is known is reasonable because, from the practical viewpoint, it is simply taken as a ``sufficiently large" number to make the $\tanh(\cdot)$ function qualify as a suitable smooth approximation of a relay.} 
%
\subsection{Two adaptive speed observers}
\lab{sec32}
%
Two solutions to the  problem presented above have been reported in \cite{ROMetal} and \cite{DAVFRIPOZ}, both are based on speed observers, that we present below.
 
\subsubsection{I\&I adaptive observer \cite[Proposition 1]{ROMetal}}
\lab{subsubsec321}
\begin{eqnarray}
			\dot x_{2I} &=& -(\hat \theta_1+k_1) \hat x_2 -\hat \theta_2 \tanh (\vartheta \hat x_2) + u,  \nonumber \\
	\hat x_2 &=& x_{2I} +k_1 x_1,  \nonumber \\	
			\dot \theta_{1I} &=& \frac{\vartheta}{k_1} \hat x_2 (\dot x_{2I} +k_1 \hat x_2 ),  \nonumber \\
	\hat \theta_1 &=&  \theta_{1I} -\frac{\vartheta}{2 k_1} \hat x^2_2, \nonumber \\
			\dot \theta_{2I} &=& \frac{\vartheta}{k_1} \tanh( \vartheta \hat x_2) (\dot x_{2I} +k_1 \hat x_2 ),  \nonumber \\
	\hat \theta_2 &=& \; \theta_{2I} -\frac{1}{k_1} \log(\cosh(\vartheta \hat x_2)), 
		\lab{iiobs}
\end{eqnarray}
where $k_1>0$ is a {\em tuning} parameter. See Appendix \ref{appa} for a discussion on the selection  of this gain.
\subsubsection{Sliding mode observer \cite[Subsection 6.2]{DAVFRIPOZ}}
\lab{subsec312}

\begin{eqnarray}
\dot{\hat{x}}_1 &=&\hat x_2+c_2 |\tilde{x}_1|^{\frac{1}{2}} \sign(\tilde{x}_1), \nonumber\\
\dot{\hat{x}}_2 &=&u+\varphi^\top \bar{\theta} +c_1 \sign(\tilde{x}_1),  \nonumber\\
\dot{\hat {\Delta}}_{\theta} &=&\Gamma \varphi[-\varphi^\top \hat {\Delta}_{\theta}+c_1 \sign(\tilde{x}_1) ], \nonumber\\
\dot{\Gamma}&=&-\Gamma\varphi\varphi^\top\Gamma,
\label{smobs}
\end{eqnarray}
where 
$$
\varphi:=\begmat{-\hat{x}_2 \\ -\tanh(\vartheta {x}_2)}
$$
with $\bar{\theta}:=[\bar \theta_1,\bar \theta_2]^\top$ is a vector of nominal values of the unknown parameters vector $\theta=[\theta_1,\theta_2]^\top$, $c_1>0$ and $c_2>0$ are {\em tuning} gains.

\subsubsection*{Comments}
\begenu[{\bf C1}]
\item Notice that we are {\em assuming known} the sign of $x_2$. Moreover, as usual in SM designs, it is assumed in  \cite{DAVFRIPOZ} that position, velocity and acceleration are {\em bounded}.
\item Denoting $\Delta_{\theta}:=\theta-\bar \theta$ the meaning of $\hat \Delta_{\theta}$ is an estimate of the error between the parameters $\theta$ and its constant a-priori estimate $\bar \theta$. Consequently, the estimates of the parameters $\theta$ are given as
\begequ
\lab{hatthe}
\hat \theta:=\hat \Delta_\theta+ \bar \theta.
\endequ
\item The matrix $\Gamma$ plays the role of the covariance matrix of the classical recursive least squares identification method with $\varphi$ representing a regressor.
\endenu
%
\subsection{Adaptive tracking controller}
\lab{subsec33}
%
Given a bounded, differentiable position reference $r(t) \in \rea$, it is easy to see that, if $x_2$ is {\em measurable} and the friction parameters are {\em known}, the ideal control law
\begequ
\lab{uid}
u^\star=\theta_1 x_2 + \theta_2 \tanh(\vartheta x_2)+\ddot r -\alpha_1 e_1 - \alpha_2 (x_2 - \dot r),
\endequ
ensures the globally convergent closed-loop dynamics 
\begalis{
\dot e_1 &= e_2\\
\dot e_2 &= -\alpha_1 e_1 - \alpha_2 e_2,
}
where $e_1:=x_1-r,e_2:=x_2-\dot r$ and $\alpha_1>0$, $\alpha_2>0$ are {\em tuning gains}. A certainty equivalent version of the control law \eqref{uid} is given by 
\begequ
\lab{u}
u=\hat \theta_1 \hat x_2 + \hat  \theta_2 \tanh(\vartheta \hat x_2)+\ddot r -\alpha_1 e_1 - \alpha_2 (\hat x_2 - \dot r),
\endequ
yielding the closed-loop dynamics of the form
\begalis{
\dot e_1 &= e_2 +\et\\
\dot e_2 &= -\alpha_1 e_1 - \alpha_2 e_2 + \et,
}
where 
\begin{equation} \label{epst}
\begin{aligned}
\et := & \theta_1 \tilde x_2 + \tilde \theta_1(x_2+\tilde x_2) + \tilde \theta_2 \tanh(\vartheta (x_2+\tilde x_2)) \\
& + \theta_2 \big[\tanh(\vartheta (x_2+\tilde x_2)) - \tanh(\vartheta x_2)\big] + \alpha_2 \tilde x_2.
\end{aligned}
\end{equation}
It is clear from \eqref{epst} that if the estimated parameters $\hat \theta_i$ and the speed estimate $\hat x_2$ converge to their true values we have that $\liminf \et(t)=0$---achieving the control objective. 

The control signal \eqref{u} was used in the simulation of both schemes, with the estimates of $x_2$ and $\hat \theta$ generated by the observers \eqref{iiobs} and \eqref{smobs}, respectively, taking into account that for the latter we use \eqref{hatthe}. 
%
\subsection{Simulation results}
\lab{sec34}
%
We select as desired trajectory $r(t)$ the curve shown in Fig. 1, which consists of an initial step plus a delayed ramp at $t=90s$:
\[
r(t) = 
\begin{cases}
1, & t < 50, \\
1.5, & 50 \leq t < 90, \\
1.5 + (0.5 - 1.5) \cdot \dfrac{t - 90}{20}, & 90 \leq t < 110, \\
0.5, & t \geq 110.
\end{cases}
\]
The tuning gains for the desired dynamics were selected as {$\alpha_1 = 0.49, \alpha_2 = 1.4$}, which corresponds to placing the two poles at $p_i=0.7$.  The parameters of the friction force were selected as {$\theta_1=0.4,\theta_2=1$ and $\vartheta=100$.}  The initial values of the system states were set as  $x_1(0)=0.1$, $x_2(0)=0.5$. 

We consider two simulation scenaria: (i) ideal measurement of all signals; and (ii) adding {\em noise} to the position measurement. We generated the additive noise by multiplying the position signal by  { $3 \times 10^{-4}$} times a random number taking values in the interval $[-1, 1]$, which corresponds to a very {\em low level} of disturbance.

In all simulations we present the following curves: 
\begenu[{\bf F1}]
\item (a) $r(t),x_1(t)$ and (b) $x_1(t)-r(t)$.
\item (a) $x_2(t),\hat x_2(t)$ and (b) $\hat x_2(t)-x_2(t)$.
\item (a) $u(t)$ and (b) $u(t)-u^\star(t)$.
\item (a) $\tilde \theta_1(t)$ and (b) $\tilde \theta_2(t)$.
\endenu
\subsubsection{I\&I observer-based controller}
\lab{subsubsec341}
%
The initial conditions for the state of this observer were taken as $x_{2I}(0)=\theta_{1I}(0)=\theta_{2I}(0)=0$ and the tuning parameter set to $k_1=1$. The explanation for the choice of this small value is given in Appendix \ref{appa}.

The simulation results for the noise-free case are shown in Figs. 1-4, while those with noise levels of $3\times 10^{-4}$ are shown in Figs. 5-8. Regardless of the presence of noise, Figs. 1 and 5 show that the state $x_1$ successfully tracks the desired trajectory $r$, and Figs. 2 and 6 show that the estimated state $\hat{x}_2$ tracks the actual state $x_2$ almost perfectly, which indicates that the proposed I \& I observer method remains effective and reliable under measurement noise. Figs. 3 and 7 illustrate that the control input $u$ is not very sensitive to noise, except at the beginning of the transient.  It should be noted from Figs. 4(a) and 8(a) that the estimated value $\hat \theta_1$ {\em does not} converge to its true value. This is due to the fact that the, relatively smooth, desired trajectory does not provide the excitation required by the theoretical analysis  \cite[Proposition 2]{ROMetal}. 
\begin{figure}[H]
\centering
\begin{subfigure}[b]{18.0pc}
    \includegraphics[width=\linewidth]{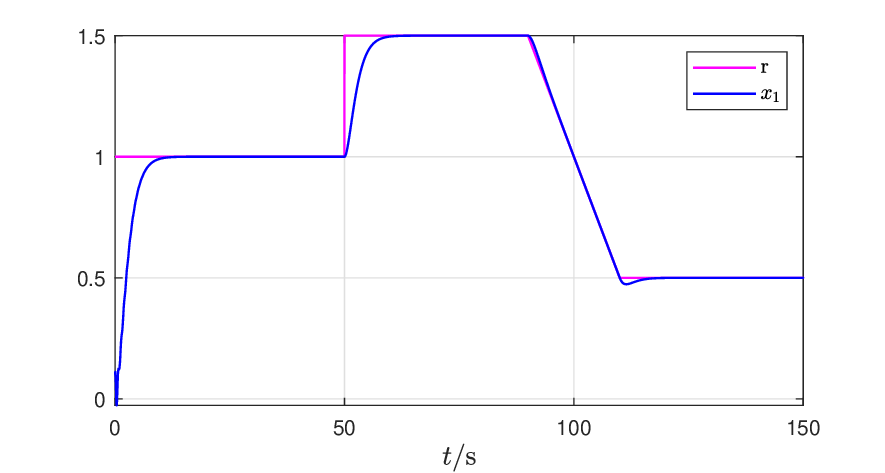}
    \caption{Trajectories of $r(t)$ and the state $x_1(t)$}
    \label{fig1a}
\end{subfigure}
\hfill
\begin{subfigure}[b]{18.0pc}
    \includegraphics[width=\linewidth]{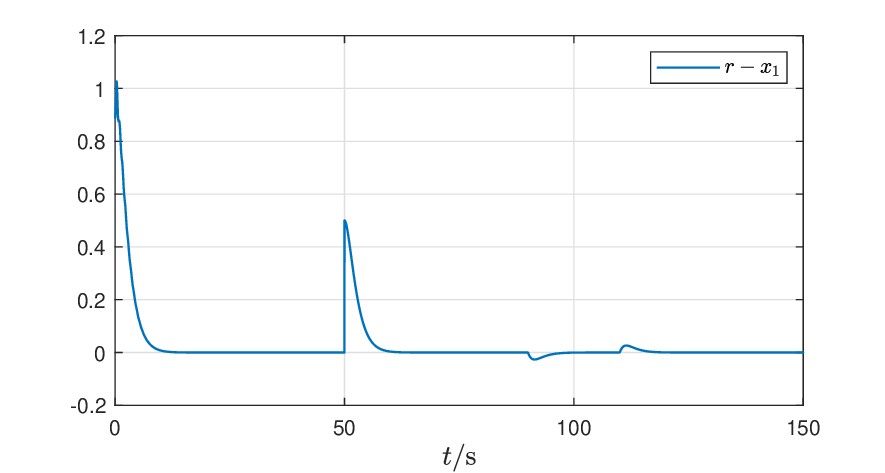}
    \caption{Trajectory of the tracking error $r(t) - x_1(t)$}
    \label{fig1b}
\end{subfigure}
\caption{Position tracking results }
\label{fig1}
\end{figure}

\begin{figure}[H]
\centering
\begin{subfigure}[b]{18.0pc}
    \includegraphics[width=\linewidth]{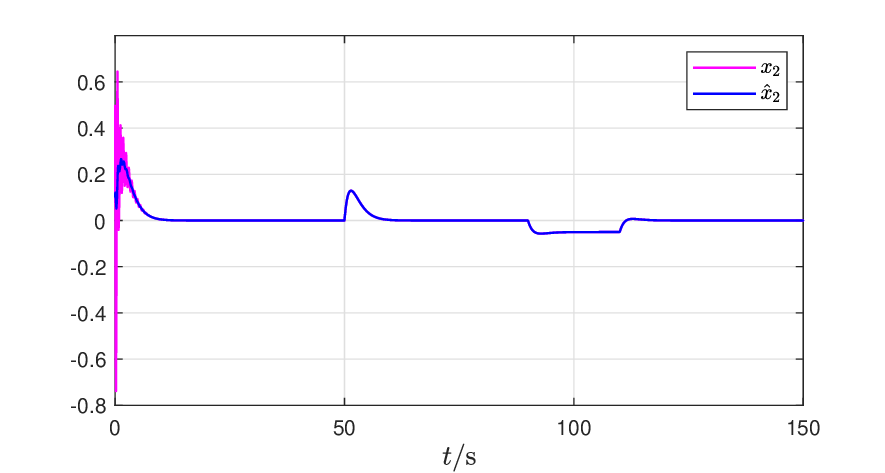}
    \caption*{(a) Trajectories of  $x_2(t)$ and its estimate $\hat{x}_2(t)$}
    \label{fig2a}
\end{subfigure}
\hfill
\begin{subfigure}[b]{18.0pc}
    \includegraphics[width=\linewidth]{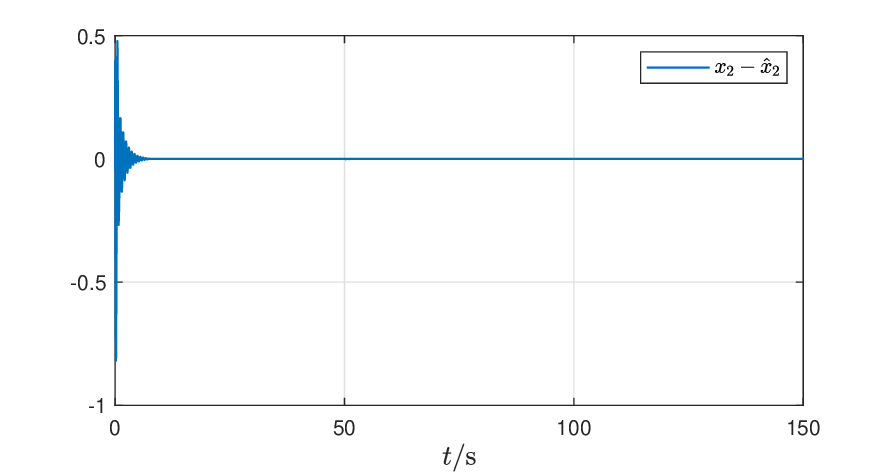}
    \caption*{(b) Trajectory of the observer error $x_2(t) - \hat{x}_2(t)$}
    \label{fig2b}
\end{subfigure}
\caption{Observer results }
\label{fig2}
\end{figure}

\begin{figure}[H]
\centering
\begin{subfigure}[b]{18.0pc}
    \includegraphics[width=\linewidth]{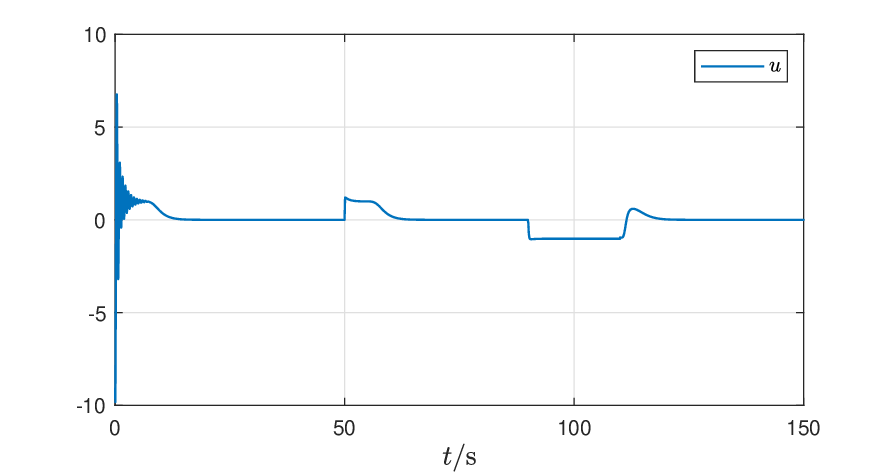}
    \caption*{(a) Trajectory of the actual control law $u(t)$}
    \label{fig3a}
\end{subfigure}
\hfill
\begin{subfigure}[b]{18.0pc}
    \includegraphics[width=\linewidth]{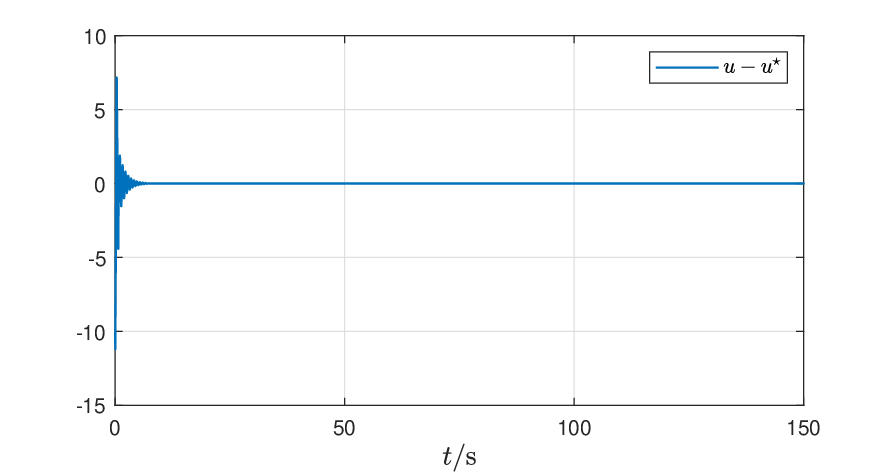}
    \caption*{(b) Trajectory of the error $u(t)- u^\star(t)$}
    \label{fig3b}
\end{subfigure}
\caption{Control signals }
\label{fig3}
\end{figure}

\begin{figure}[H]
\centering
\begin{subfigure}[b]{18.0pc}
    \includegraphics[width=\linewidth]{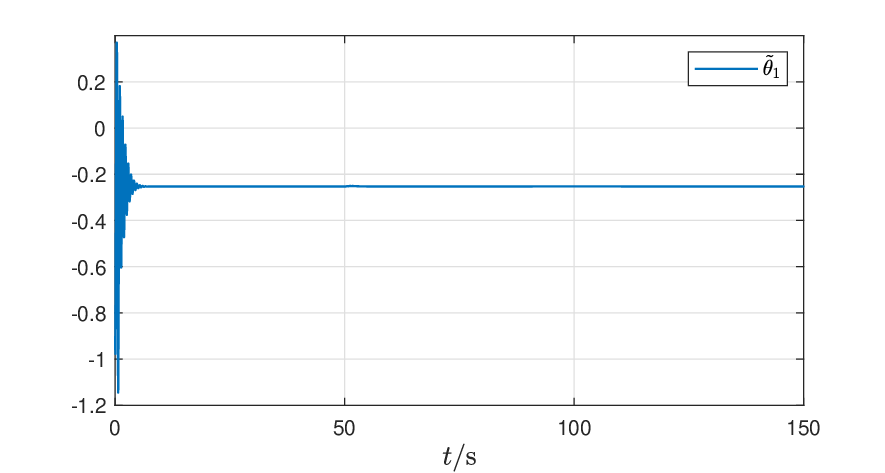}
    \caption*{(a) Trajectory of the parameter estimate error $\tilde \theta_1(t)$}   
    \label{fig4a}
\end{subfigure}
\hfill
\begin{subfigure}[b]{18.0pc}
    \includegraphics[width=\linewidth]{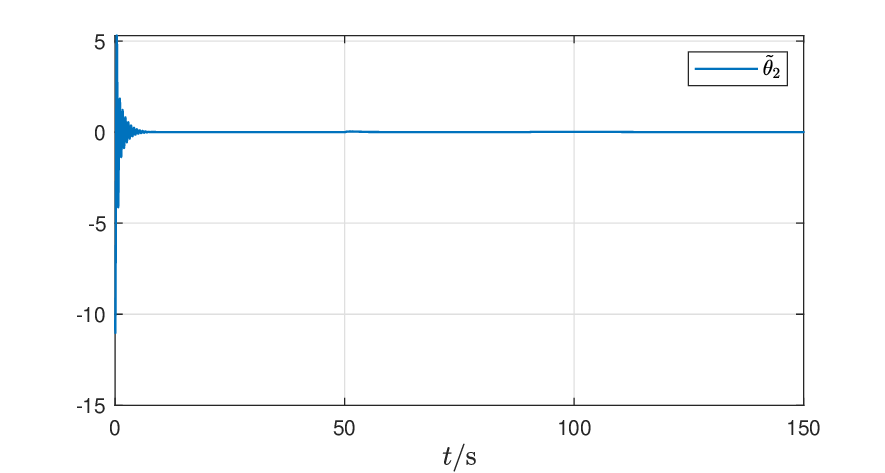}
    \caption*{(b) Trajectory of the parameter estimate error $\tilde \theta_2(t)$ }
    \label{fig4b}
\end{subfigure}
\caption{Parameter estimation errors}
\label{fig4}
\end{figure}

\begin{figure}[H]
\centering
\begin{subfigure}[b]{18.0pc}
    \includegraphics[width=\linewidth]{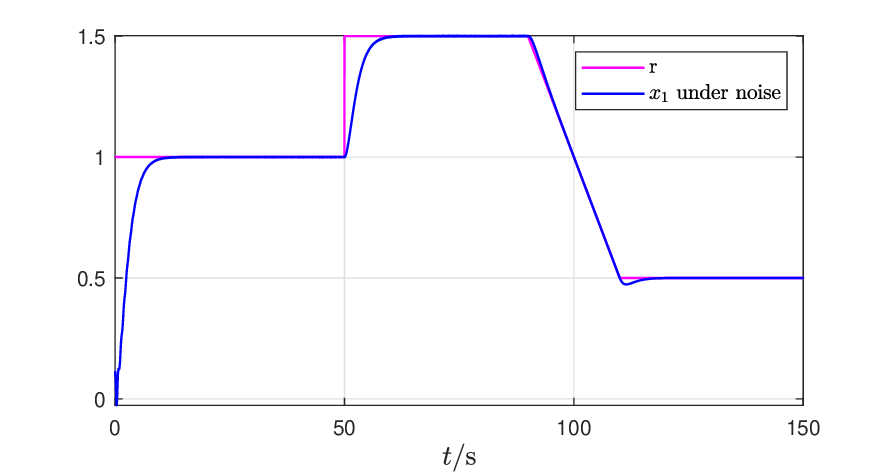}
    \caption{Trajectories of $r(t)$ and the state $x_1(t)$: noisy case}
    \label{fig5a}
\end{subfigure}
\hfill
\begin{subfigure}[b]{18.0pc}
    \includegraphics[width=\linewidth]{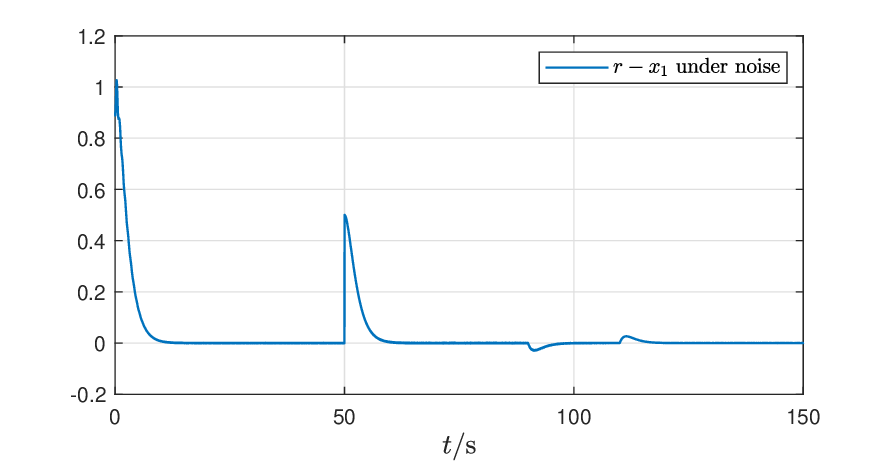}
    \caption{Trajectory of the tracking error $r(t) - x_1(t)$: noisy case }
    \label{fig5b}
\end{subfigure}
\caption{Position tracking results: noisy case}
\label{fig5}
\end{figure}

\begin{figure}[H]
\centering
\begin{subfigure}[b]{18.0pc}
    \includegraphics[width=\linewidth]{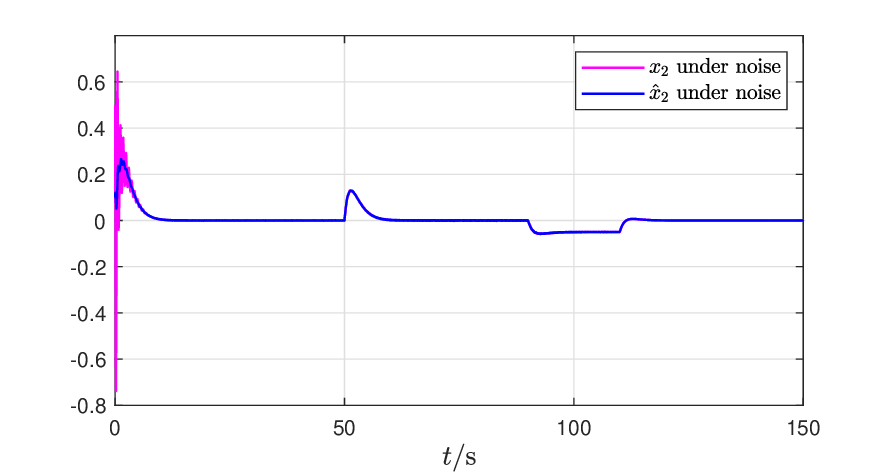}
    \caption*{(a) Trajectories of  $x_2(t)$ and its estimate $\hat{x}_2(t)$: noisy case}
    \label{fig6a}
\end{subfigure}
\hfill
\begin{subfigure}[b]{18.0pc}
    \includegraphics[width=\linewidth]{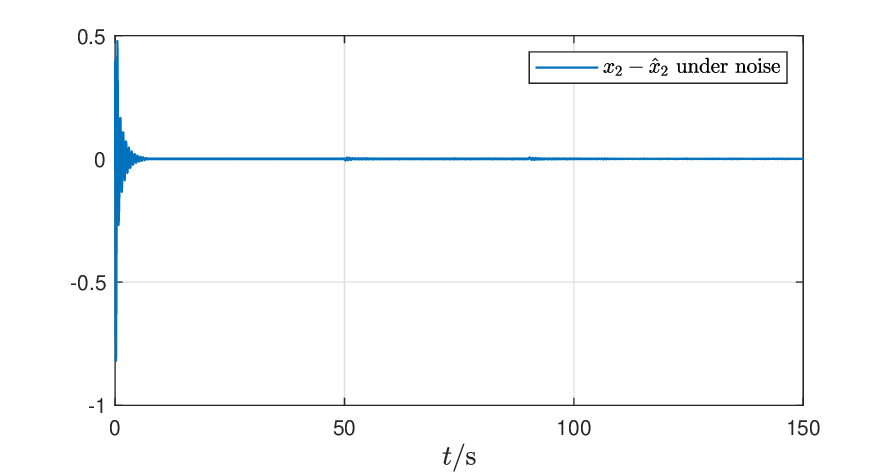}
    \caption*{(b) Trajectory of the observer error $x_2(t) - \hat{x}_2(t)$: noisy case}
    \label{fig6b}
\end{subfigure}
\caption{Observer results: noisy case }
\label{fig6}
\end{figure}

\begin{figure}[H]
\centering
\begin{subfigure}[b]{18.0pc}
    \includegraphics[width=\linewidth]{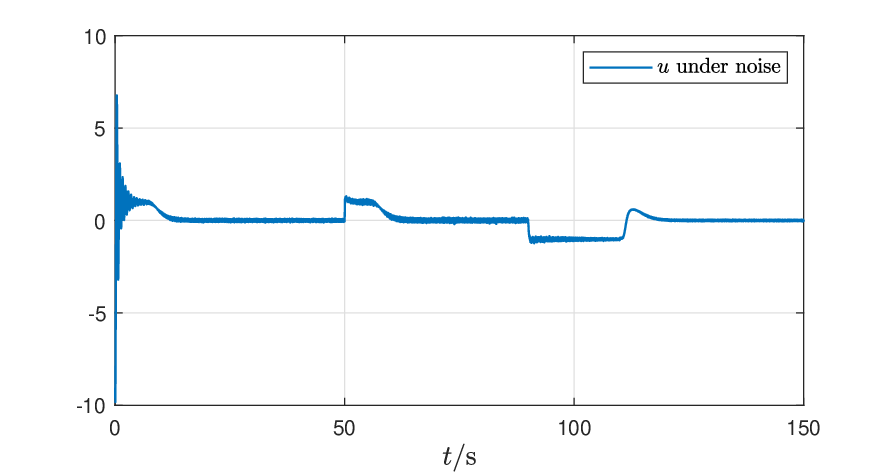}
    \caption*{(a) Trajectory of the actual control law $u(t)$: noisy case}
    \label{fig7a}
\end{subfigure}
\hfill
\begin{subfigure}[b]{18.0pc}
    \includegraphics[width=\linewidth]{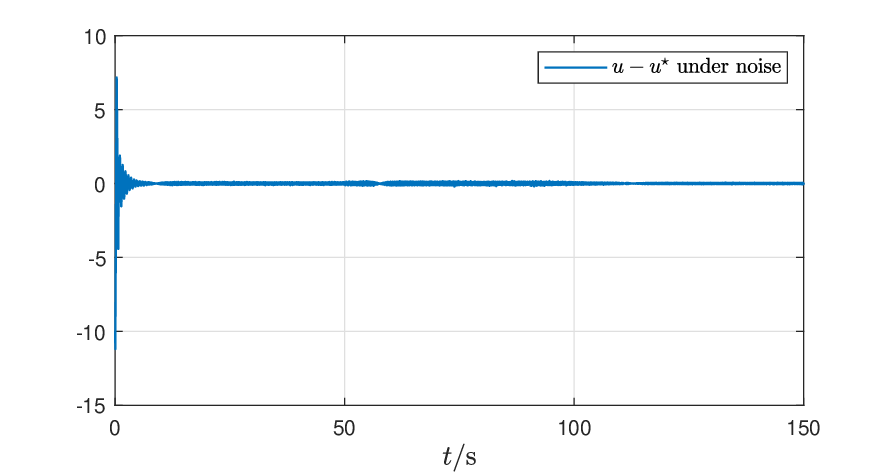}
    \caption*{(b) Trajectory of the error $u(t)- u^\star(t)$: noisy case}
    \label{fig7b}
\end{subfigure}
\caption{Control signals: noisy case}
\label{fig7}
\end{figure}

\begin{figure}[H]
\centering
\begin{subfigure}[b]{18.0pc}
    \includegraphics[width=\linewidth]{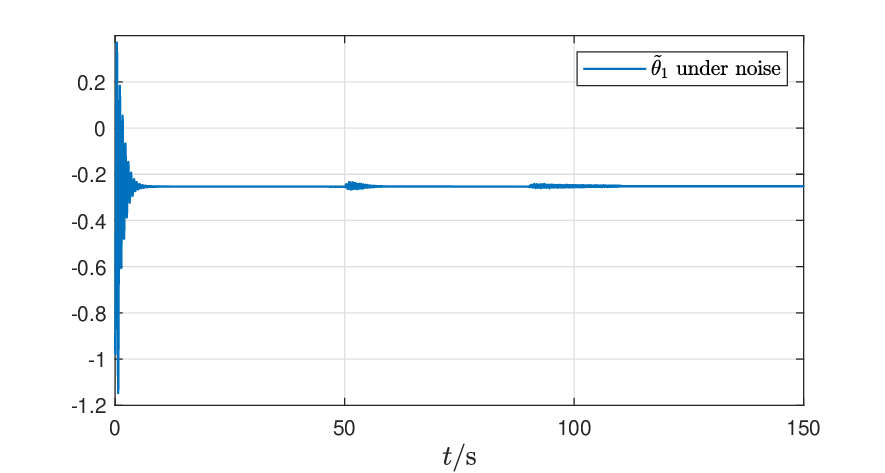}
    \caption*{(a) Trajectory of the parameter estimate error $\tilde \theta_1(t)$: noisy case}
    \label{fig8a}
\end{subfigure}
\hfill
\begin{subfigure}[b]{18.0pc}
    \includegraphics[width=\linewidth]{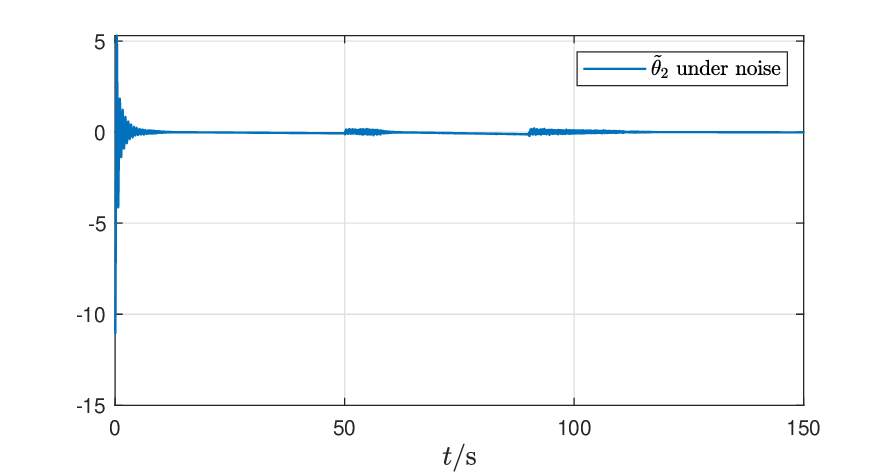}
    \caption*{(b) Trajectory of the parameter estimate error $\tilde \theta_2(t)$: noisy case}
    \label{fig8b}
\end{subfigure}
\caption{Parameter estimation errors: noisy case}
\label{fig8}
\end{figure}

\subsubsection{Sliding mode observer-based controller}
\lab{subsubsec342}

The initial conditions for the state of the SM observer were taken to be the same as those of the I\&I observer, specifically:  
\begalis{
\hat x_1(0) &=0,\\
 \hat x_{2}(0) & =x_{2I}(0)+k_1 x_1(0)=0.1,\\
 \bar \theta &=\begmat{0.5\theta_{1}\\ 0.5\theta_{2}}=\begmat{0.2\\ 0.5},\\
\hat {\Delta}_{\theta} & =\begmat{\hat \theta_1(0)\\ \hat \theta_2(0)}-\bar \theta, \\
\Gamma(0) &=\diag \{500,500\},
}
and the tuning parameter set to $c_1=0.5$ and $c_2=25$---whose selection implied an exhaustive and hard trial and error process. The initial conditions for the parameter estimates was done such that we get the same values we have for the I\&I observer, namely:
\begalis{
\hat \theta_{1}(0) &=\theta_{1I}(0) -\frac{\vartheta}{2 k_1} \hat x^2_2(0)\\
\hat \theta_{2}(0) &=\theta_{2I}(0) -\frac{1}{k_1} \log(\cosh(\vartheta \hat x_2(0))).
}

The simulation results for the noise-free case are shown in Figs. 9-12, while the noisy one are shown in Figs. 13-16. {Fig. 9(a) shows that, in the noise-free case, the state $x_1$ follows the desired trajectory $r$ while exhibiting a slight oscillatory behavior. It should be pointed out that, as seen in Fig. 9(b), that estimation of $x_1$ is very accurate, with the blue and the green line overlaping. However, as shown in Fig. 13,  the presence of noise causes $x_1$ and $\hat x_1$ to diverge---clearly indicating that the closed-loop system is highly sensitive to small levels of measurement noise. It can be clearly observed from Fig. 10  that, without measurement noise, the SM observer estimate $\hat{x}_2$ tracks the actual state $x_2$ almost perfectly. In contrast, from Fig. 14 we see that, when measurement noise is present, the estimated state $\hat{x}_2$ fails to converge and exhibits steady-state error, showing again that the SM observer is highly sensitive to measurement noise. From Fig. 11 it can be seen that, even in the noise-free case, $u$ exhibits an oscillatory behavior. This is because the control signal $u$ contains a term $\hat{\theta}_2 \tanh(\vartheta \hat{x}_2)$ with  $\vartheta = 100$ and the estimated state $\hat{x}_2$ oscillates around zero even without noise. {It should be noted from Figs. 12(a) and 16 that, in the noise-free case, the estimated value $\hat{\theta}_1$, and in the noisy case, both $\hat{\theta}_1$ and $\hat{\theta}_2$, {\em do not} converge to their true value. This is due to the fact that the relatively smooth desired trajectory, which fails to provide sufficient excitation.}  

\begin{figure}[H]
\centering
\begin{subfigure}[b]{18.0pc}
    \includegraphics[width=\linewidth]{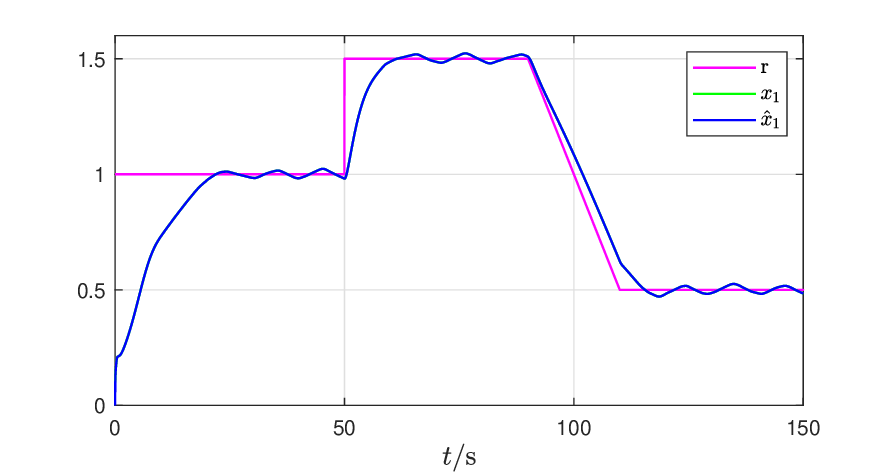}
    \caption{Trajectories of $r(t)$, the state $x_1(t)$ and its estimate}
\end{subfigure}
\hfill
\begin{subfigure}[b]{18.0pc}
    \includegraphics[width=\linewidth]{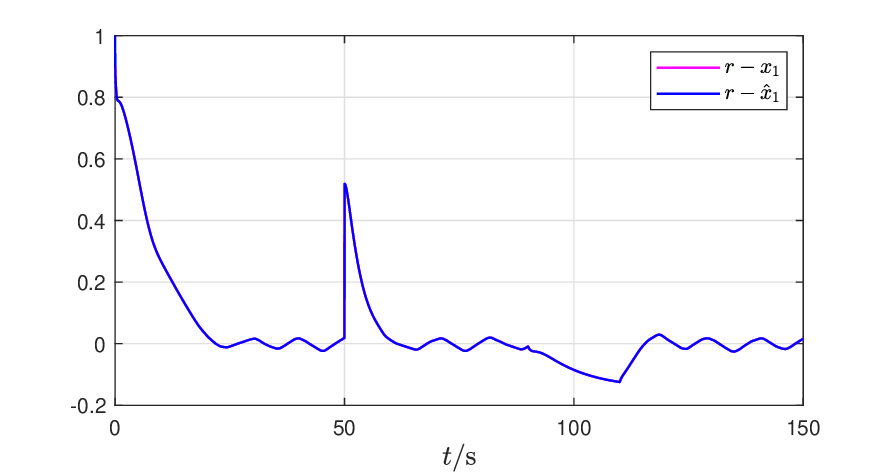}
    \caption{Trajectory of $r(t) - x_1(t)$ and $r(t) - \hat x_1(t)$}
\end{subfigure}
\caption{Position tracking results}
\label{fig1}
\end{figure}

\begin{figure}[H]
\centering
\begin{subfigure}[b]{18.0pc}
    \includegraphics[width=\linewidth]{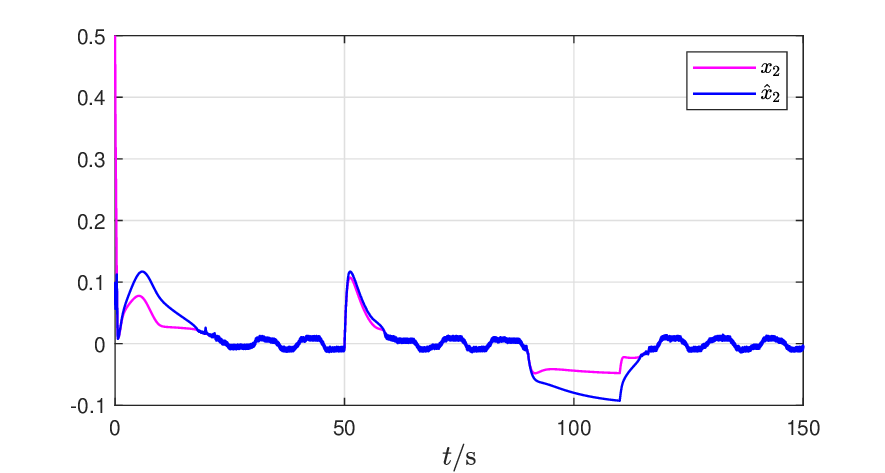}
    \caption*{(a) Trajectories of  $x_2(t)$ and its estimate $\hat{x}_2(t)$}
\end{subfigure}
\hfill
\begin{subfigure}[b]{18.0pc}
    \includegraphics[width=\linewidth]{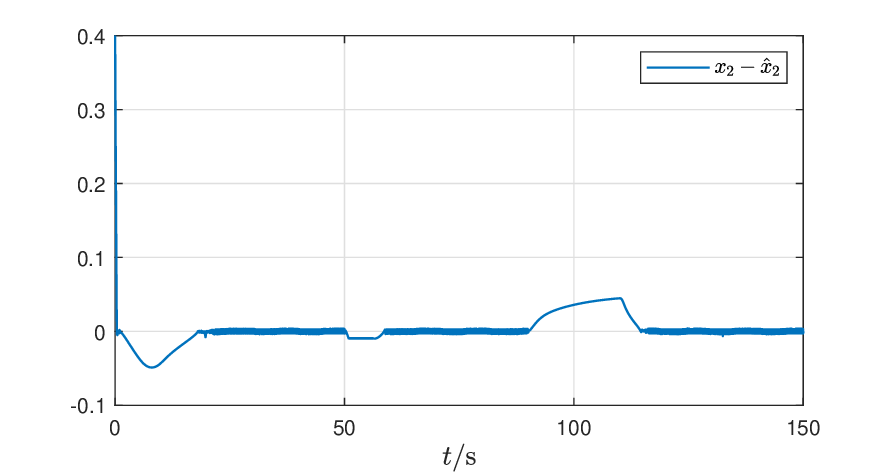}
    \caption*{(b) Trajectory of the observer error $x_2(t) - \hat{x}_2(t)$}
\end{subfigure}
\caption{Observer results }
\label{fig2}
\end{figure}

\begin{figure}[H]
\centering
\begin{subfigure}[b]{18.0pc}
    \includegraphics[width=\linewidth]{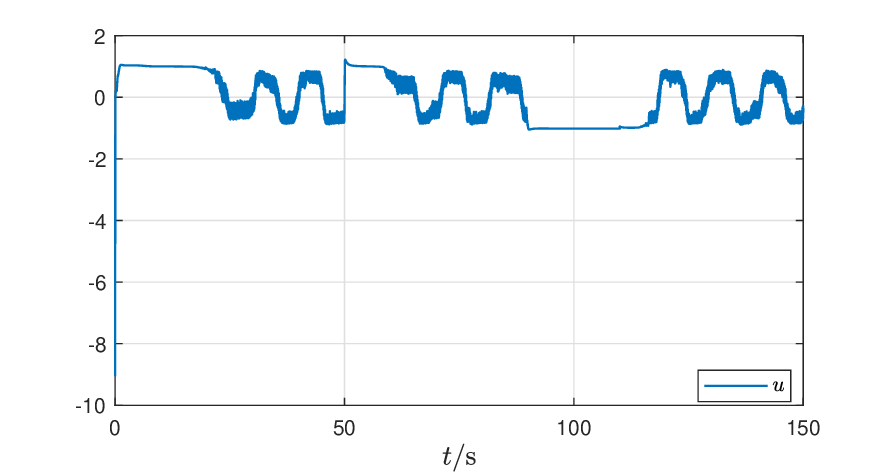}
    \caption*{(a) Trajectory of the actual control law $u(t)$}
\end{subfigure}
\hfill
\begin{subfigure}[b]{18.0pc}
    \includegraphics[width=\linewidth]{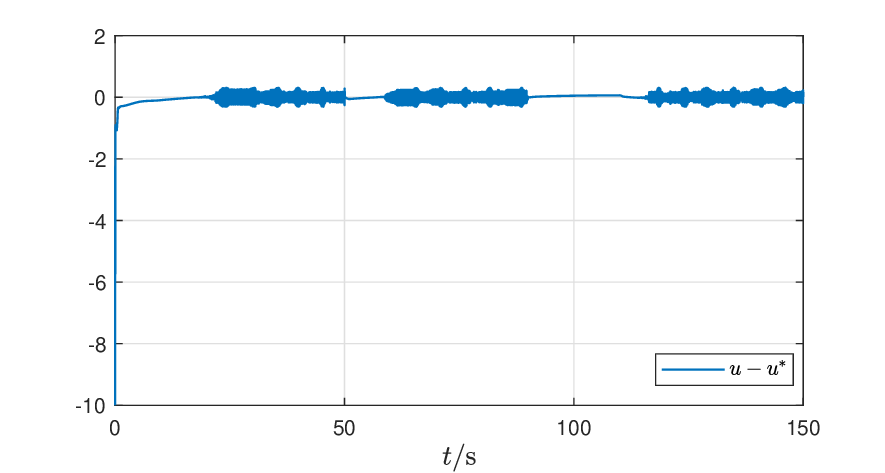}
    \caption*{(b) Trajectory of the error $u(t)- u^\star(t)$}
\end{subfigure}
\caption{Control signals}
\label{fig7}
\end{figure}

\begin{figure}[H]
\centering
\begin{subfigure}[b]{18.0pc}
    \includegraphics[width=\linewidth]{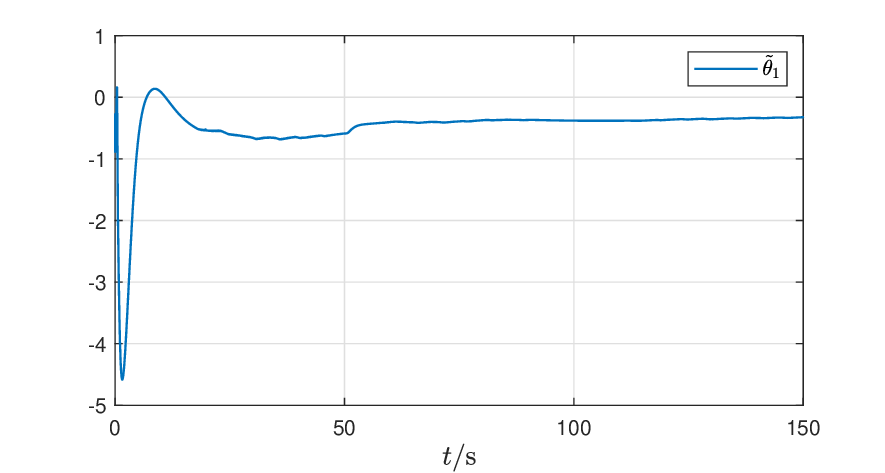}
    \caption*{(a) Trajectory of the parameter estimate error $\tilde \theta_1(t)$}
\end{subfigure}
\hfill
\begin{subfigure}[b]{18.0pc}
    \includegraphics[width=\linewidth]{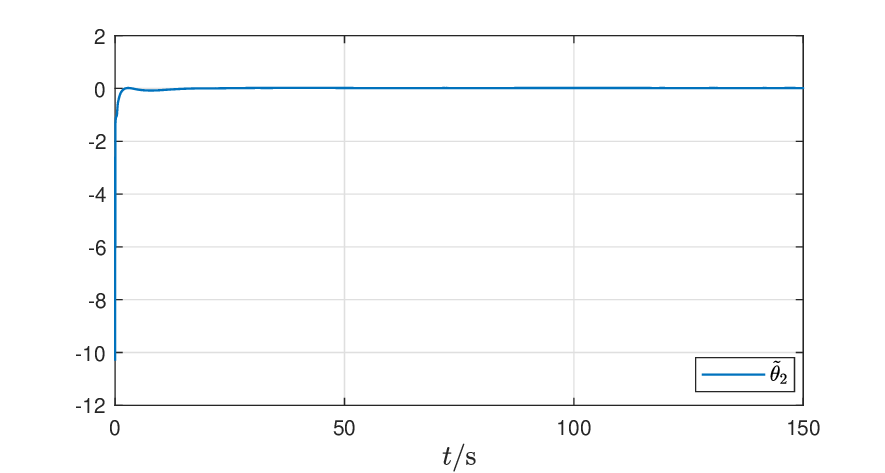}
    \caption*{(b) Trajectory of the parameter estimate error $\tilde \theta_2(t)$}
\end{subfigure}
\caption{Parameter estimation errors}
\label{fig4}
\end{figure}

\begin{figure}[H]
\centering
\begin{subfigure}[b]{18.0pc}
    \includegraphics[width=\linewidth]{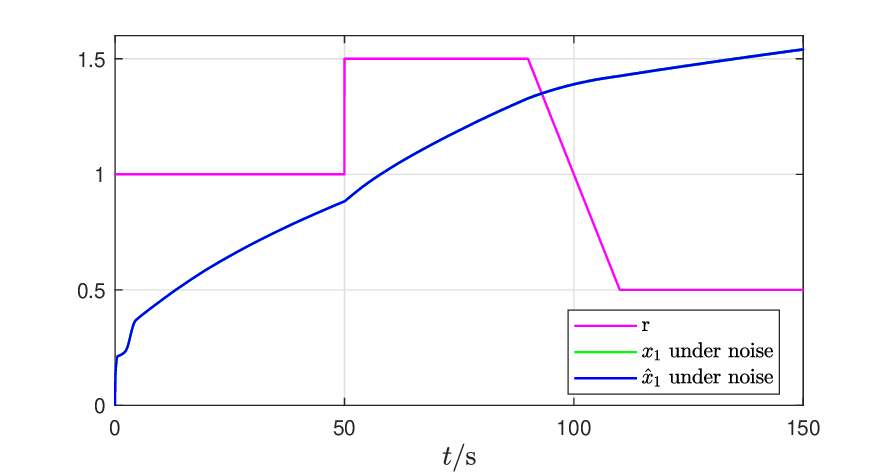}
    \caption{Trajectories of $r(t)$, the state $x_1(t)$ and its estimate: noisy case }
\end{subfigure}
\hfill
\begin{subfigure}[b]{18.0pc}
    \includegraphics[width=\linewidth]{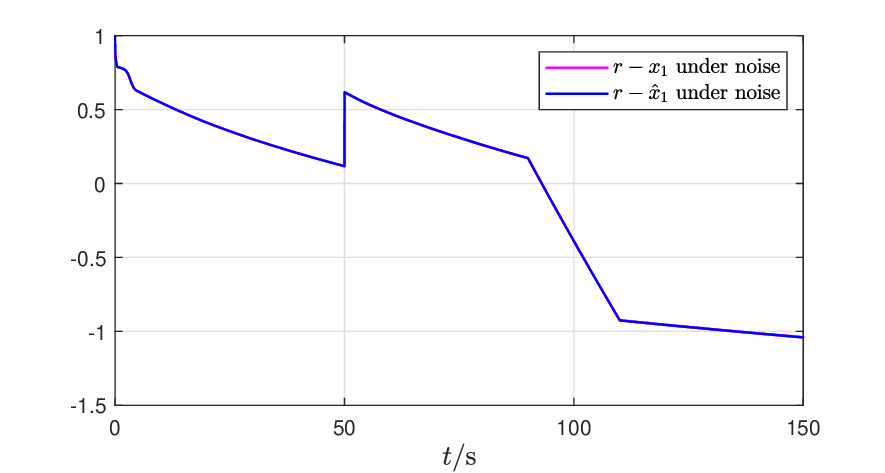}
    \caption{Trajectory of $r(t) - x_1(t)$ and $r(t) - \hat x_1(t)$: noisy case }
\end{subfigure}
\caption{Position tracking results: noisy case}
\label{fig5}
\end{figure}

\begin{figure}[H]
\centering
\begin{subfigure}[b]{18.0pc}
    \includegraphics[width=\linewidth]{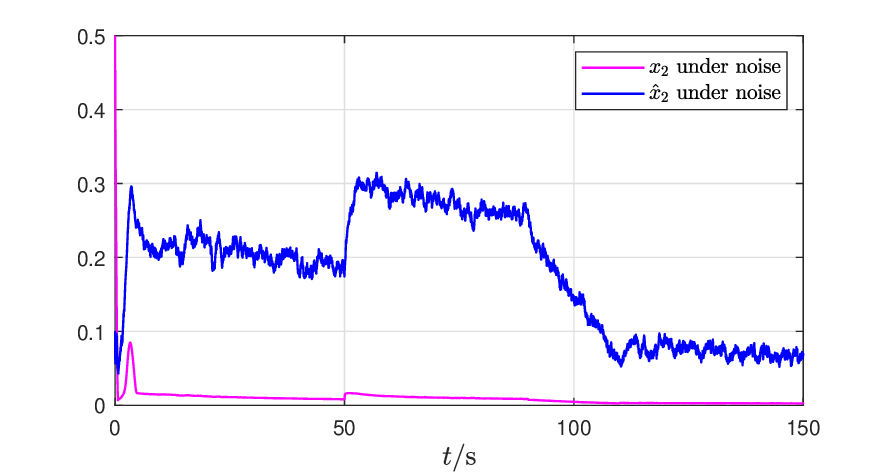}
    \caption*{(a) Trajectories of  $x_2(t)$ and its estimate $\hat{x}_2(t)$: noisy case }
\end{subfigure}
\hfill
\begin{subfigure}[b]{18.0pc}
    \includegraphics[width=\linewidth]{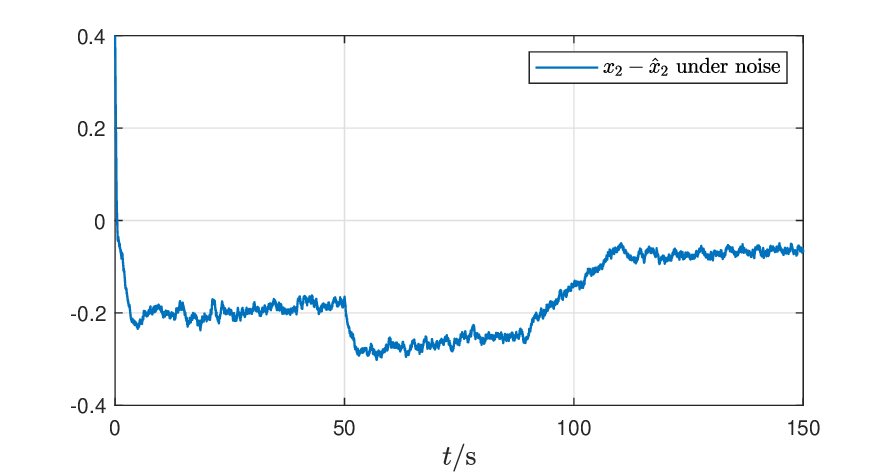}
    \caption*{(b) Trajectory of the observer error $x_2(t) - \hat{x}_2(t)$: noisy case }
\end{subfigure}
\caption{Observer results: noisy case}
\label{fig6}
\end{figure}

\begin{figure}[H]
\centering
\begin{subfigure}[b]{18.0pc}
    \includegraphics[width=\linewidth]{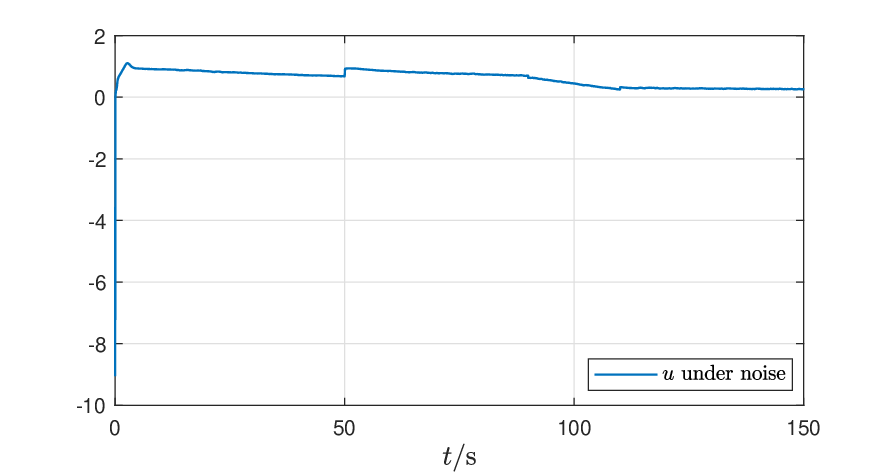}
    \caption*{(a) Trajectory of the actual control law $u(t)$: noisy case }
\end{subfigure}
\hfill
\begin{subfigure}[b]{18.0pc}
    \includegraphics[width=\linewidth]{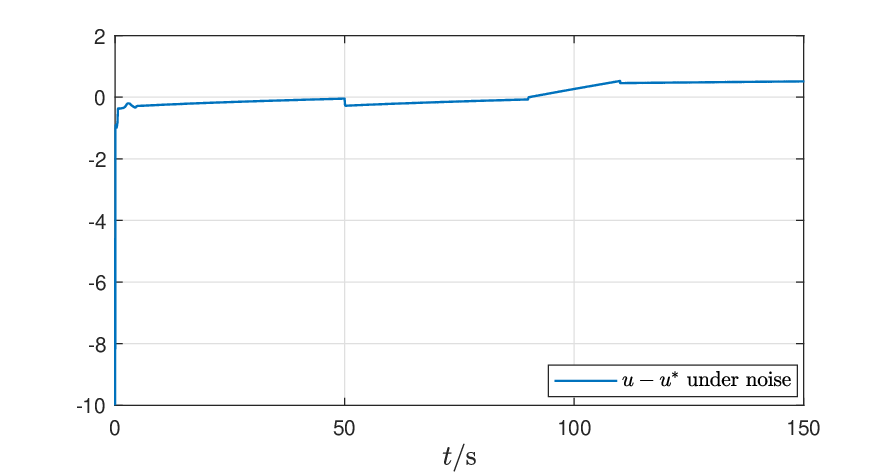}
    \caption*{(b) Trajectory of the error $u(t)- u^\star(t)$: noisy case }
\end{subfigure}
\caption{Control signals: noisy case }
\label{fig7}
\end{figure}

\begin{figure}[H]
\centering
\begin{subfigure}[b]{18.0pc}
    \includegraphics[width=\linewidth]{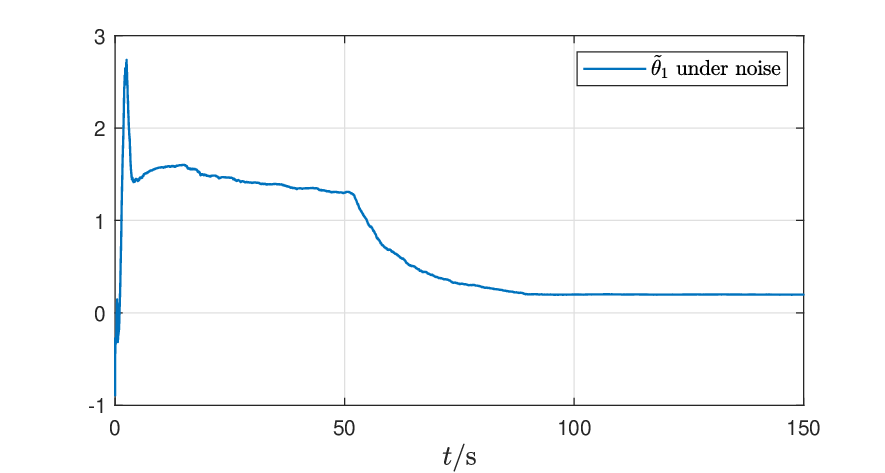}
    \caption*{(a) Trajectory of the parameter estimation error $\tilde \theta_1(t)$: noisy case }
\end{subfigure}
\hfill
\begin{subfigure}[b]{18.0pc}
    \includegraphics[width=\linewidth]{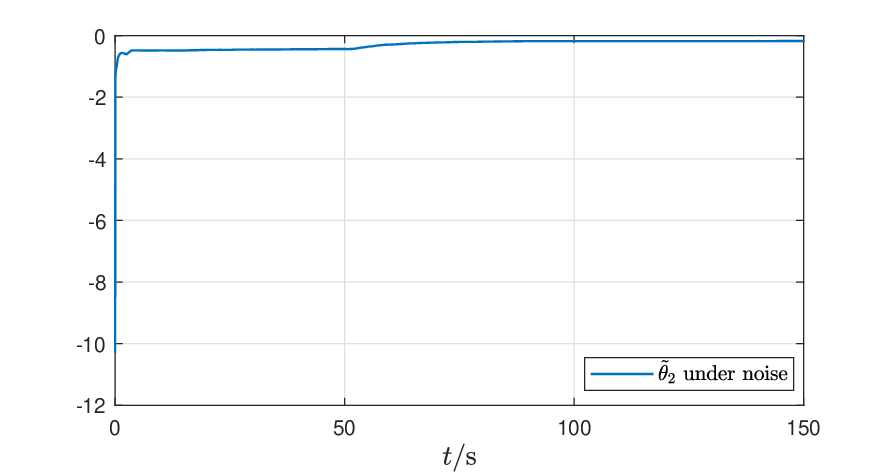}
    \caption*{(b) Trajectory of the parameter estimate error $\tilde \theta_2(t)$: noisy case }
\end{subfigure}
\caption{Parameter estimation errors: noisy case }
\label{fig8}
\end{figure}
\section{Concluding Remarks}
\lab{sec4}
%
The control designs predominantly used in engineering applications preserve, and effectively exploit, the physical structure of the system. Some examples being:
\begenu[{\bf E1}]
\item In robotic systems the dominant control strategy in regulation tasks is the well-known Takegaki-Arimoto {\em PD control}  \cite[Proposition 3.1]{ORTetalbookel}, that exploits the physical property of {\em passivity} of the map between force and velocity. For tracking applications the standard scheme is the Slotine and Li controller    \cite[Subsection 4.1.2]{ORTetalbookel}, that relies on the key physical property 
$$
\dot M(q)=C(q,\dot q)+C^\top(q,\dot q),
$$
where $M(q)$ is the inertia matrix and $C(q,\dot q)$ the matrix of Coriolis and centrifugal forces.
\item In electrical motors the industry standard is the {\em Field Oriented Control} \cite[Subsection  10.5]{ORTetalbookel}, that heavily relies on a suitable representation of the motor dynamics in a rotating reference frame. Another technique, that is widely used in motor applications, is {\em Direct Torque Control}, whose switching pattern was rigorously analyzed and improved in \cite{ORTBARESC}.
\item In power electronics one of the leading strategies is Akagi's Instantaneous Power Theory, whose rigorous mathematical formulation was established in \cite[Subsection  4.4]{ORTetalbookpid} invoking the physical property of {\em shifted passivity}. 
\item In power systems, in particular in microgrids, the universally adopted scheme is the {\em Droop Control}, studied in the award winning paper   \cite{SCHetal} invoking Lyapunov theory. 
\endenu 

Given that engineering practice is dominated by control schemes that preserve, and exploit, the systems physical structure, one is left to wonder what is the interest to study schemes whose first basic step is to destroy the structure, as done in SM? Equally important is the question: Isn't it dangerous to train the new generation of engineers in topics that don't have an impact in current practice? In the same vein, shouldn't the scientific publications in control consider imposing higher quality standards for this kind of material?\footnote{We just learned that in the forthcoming IEEE-CDC25 there will be 24 papers on SM!.} 

Let us close this discussion recalling that in the early 90's there were discussions on the IEEE-TAC Editorial Board on the pertinence of publishing papers on {\em fuzzy control}. The decision was taken to suggest the leading authors in the field to create their own journal, which turned out to be a very pertinent, and successful, initiative. 

\section{Author Information}
{
\begin{IEEEbiography}{{R}omeo Ortega}{\,}(romeo.ortega@itam.mx) received the Ph.D. degree (Docteur d'Etat) from the Polytechnical Institute of Grenoble, France. He is a full time professor at Instituto Tecnol\'ogico Aut\'onomo de M\'exico (ITAM), Mexico. His focus is on the fields of nonlinear and adaptive control, with special emphasis on application. He is IEEE Life Fellow and IFAC Fellow. 
\end{IEEEbiography}}

\begin{IEEEbiography}{Leyan Fang}{\,} received the M.S. degree from Harbin Institute of Technology, Harbin, China, where she is currently pursuing the Ph.D. degree. Her focus is on adaptive control and parameter estimation.
\end{IEEEbiography}

{
\begin{IEEEbiography}{Jose Guadalupe Romero}{\,} received the Ph.D. degree from the University of Paris-Sud XI, France. He is a full time professor at Instituto Tecnol\'ogico Aut\'onomo de M\'exico (ITAM), Mexico. His focus is on nonlinear and adaptive control, stability analysis  and the state estimation problem, with application to mechanical systems, aerial vehicles, mobile robots and multi-agent systems.
\end{IEEEbiography}}

\appendices
%
\section{Some Remarks on the Gain Tuning of the I\&I Observer}
\lab{appa}
%
It has been argued above that another shortcoming of SM designs is the proliferation of tuning gains, for whose selection no guidelines are provided. This situation significantly complicates the, already difficult, task of commissioning the controller/observer design. In counterposition to SM designs, for controllers/observers that do not destroy the structure of the system, we usually have some guidelines to select these gains. 

As a token of illustration we briefly discuss below some tuning guidelines for the gain $k_1$, which is the only free tuning gain of the proposed I\&I observer. 

Given the parameters \(\vartheta = 100, \ \theta_{1} = 0.4, \ \theta_{2} = 1\):  

\begin{enumerate}
  \item \textbf{Effect of \(k_{1}\) on the convergence rate of the Lyapunov function:} The stability analysis of the I\&I observer relies on the construction of a Lyapunov function $\calh$ whose time derivative satisfies the bound   
  \[
  \dot{{\mathcal H}} \le -\vartheta^{2} \left(k_{1} + \theta_{1} + \theta_{2} \vartheta \right) \tilde{x}_{2}^{2}.
  \]
  Substituting the number used in the simulation we get 
  \[
  \dot{{\mathcal H}} \le -100^{2} \left( k_{1}+ 100.4 \right) \tilde{x}_{2}^{2}.
  \]
  This shows that increasing \(k_{1}\) indeed improves the convergence rate of the energy function. However, since \(\theta_{1}+\theta_{2} \vartheta = 100.4\) is large, choosing a small \(k_{1}\) will not significantly affect the convergence rate of $\mathcal H$. On the other hand, to have an impact on the convergence rate we need to increasing the value of \(k_{1}\)  on the same order of magnitude of $100$. But in this case, other considerations discussed below enter into the picture.  

  \item \textbf{Effect of \(k_{1}\) on the adaptive observer:}  The I\&I observer contains a term of the form 
  \[
  \hat{x}_{2} = x_{2I} + k_{1} x_{1}.
  \]
  A larger \(k_{1}\) increases the proportion of the position \(x_{1}\) in the velocity estimate, which amplifies position measurement noise and may cause oscillations or even divergence in the estimate.  

  Meanwhile,  
  \[
  \dot{x}_{2I} = -\left( \hat{\theta}_{1} + k_{1} \right) \hat{x}_{2} - \hat{\theta}_{2} \tanh\!\left( \vartheta \hat{x}_{2} \right) + u,
  \]
  where a larger \(k_{1}\) makes the observer ``faster''. This requires a smaller integration step size in simulations to avoid numerical instability or peaking phenomena.  

  \item \textbf{Effect of \(k_{1}\) on the parameter adaptation laws:} The parameter estimation errors in the I\&I observer satisfy   
\[
\begin{aligned}
\dot{\tilde{\theta}}_1 &= -\vartheta \hat{x}_2, \\
\dot{\tilde{\theta}}_2 &= -\vartheta \tanh(\vartheta \hat{x}_2),
\end{aligned}
\]
which does not explicitly depend on \(k_{1}\). Therefore, \(k_{1}\) has little direct effect on the parameter estimation errors.
\end{enumerate}

In summary, the tuning parameter $k_1$ requires a careful balance: choosing a moderate value ensures a good convergence rate while avoiding excessive noise amplification and numerical stiffness (i.e., some state variables change very rapidly while others change much more slowly).

Simulation results show that when $k_1$ is chosen within the interval (0, 88], the state observation is stable and performs well; however, when $k_1$ exceeds this range, the sensitivity to noise significantly increases, which may cause large oscillations or even divergence in the estimation.

\endarticle

\end{document}